\documentclass[aps,prl,twocolumn,superscriptaddress]{revtex4}
\usepackage{graphics}
\usepackage{graphicx}
\usepackage{amssymb}
\usepackage{hyperref}
\usepackage{color}

\begin{document}

\title{Structural investigations on $\epsilon$-FeGe
at high pressure and low temperature}

\author{H. Wilhelm}
\affiliation{Diamond Light Source Ltd., Chilton, Didcot,
Oxfordshire, OX11 0DE, United Kingdom}
\author{M. Schmidt}
\affiliation{Max Planck Institut f\"ur Chemische Physik fester
Stoffe, N\"othnitzer Str. 40, 01187 Dresden, Germany}
\author{R. Cardoso-Gil}
\affiliation{Max Planck Institut f\"ur Chemische Physik fester
Stoffe, N\"othnitzer Str. 40, 01187 Dresden, Germany}
\author{U. Burkhardt}
\affiliation{Max Planck Institut f\"ur Chemische Physik fester
Stoffe, N\"othnitzer Str. 40, 01187 Dresden, Germany}
\author{M. Hanfland}
\affiliation{European Synchrotron Radiation Facility, 38043
Grenoble, Cedex, France}
\author{U. Schwarz}
\affiliation{Max Planck Institut f\"ur Chemische Physik fester
Stoffe, N\"othnitzer Str. 40, 01187 Dresden, Germany}
\author{L. Akselrud}
\affiliation{Ivan Franko Lviv National University, Kyrylo and
Mephodiy Street 6, 79005 Lviv, Ukraine}

\begin{abstract}
The structural parameters of $\epsilon$-FeGe have been determined at
ambient conditions using single crystal refinement. Powder
diffraction have been carried out to determine structural properties
and compressibility for pressures up to 30 GPa and temperatures as
low as 82 K. The discontinuous change in the pressure dependence of
the shortest Fe-Ge interatomic distance might be interpreted as a
symmetry-conserving transition and seems to be related to a magnetic
phase boundary line.
\end{abstract}

\maketitle

\section{Introduction}
Binary transition metal compounds crystallizing in the $B20$
structure (space group $P2_13$) \cite{Boren33,Pauling48} show a
variety of interesting physical properties ranging from a
narrow-gapped semiconductor (FeSi) to diamagnetic (CoSi) or
paramagnetic (FeGe) metal. In the latter case recent experiments
have shown that the helical ordering temperature ($T_C=280$~K at
ambient pressure) can be suppressed by the application of external
pressure \cite{Pedrazzini07}. Furthermore, anomalies in internal
structural parameters in cubic FeGe ($\epsilon$-FeGe) for isothermal
compression seem to be related to the magnetic phase boundary line.
Here we present single crystal data at ambient conditions as well as
in detail the results of the powder diffraction data at high
pressure and temperatures as low as 82~K.

\section{Sample Growth and Characterization}
Single crystals of $\epsilon$-FeGe were grown by chemical vapor
transport using iodine as agent \cite{Richardson67} in a homemade
two zone furnace. $\epsilon$-FeGe crystallized very slowly by an
endothermal transport reaction from 850~K to 810~K although the
transport was made perpendicular to the tube axis \cite{Kraemer74}.
The resulting crystals had a volume of less than 1~mm$^3$. Pieces
used in the subsequent experiments were examined thoroughly by
various X-ray techniques and electron-beam microanalysis.

\begin{figure}
\center{\includegraphics[width=0.35\textwidth]{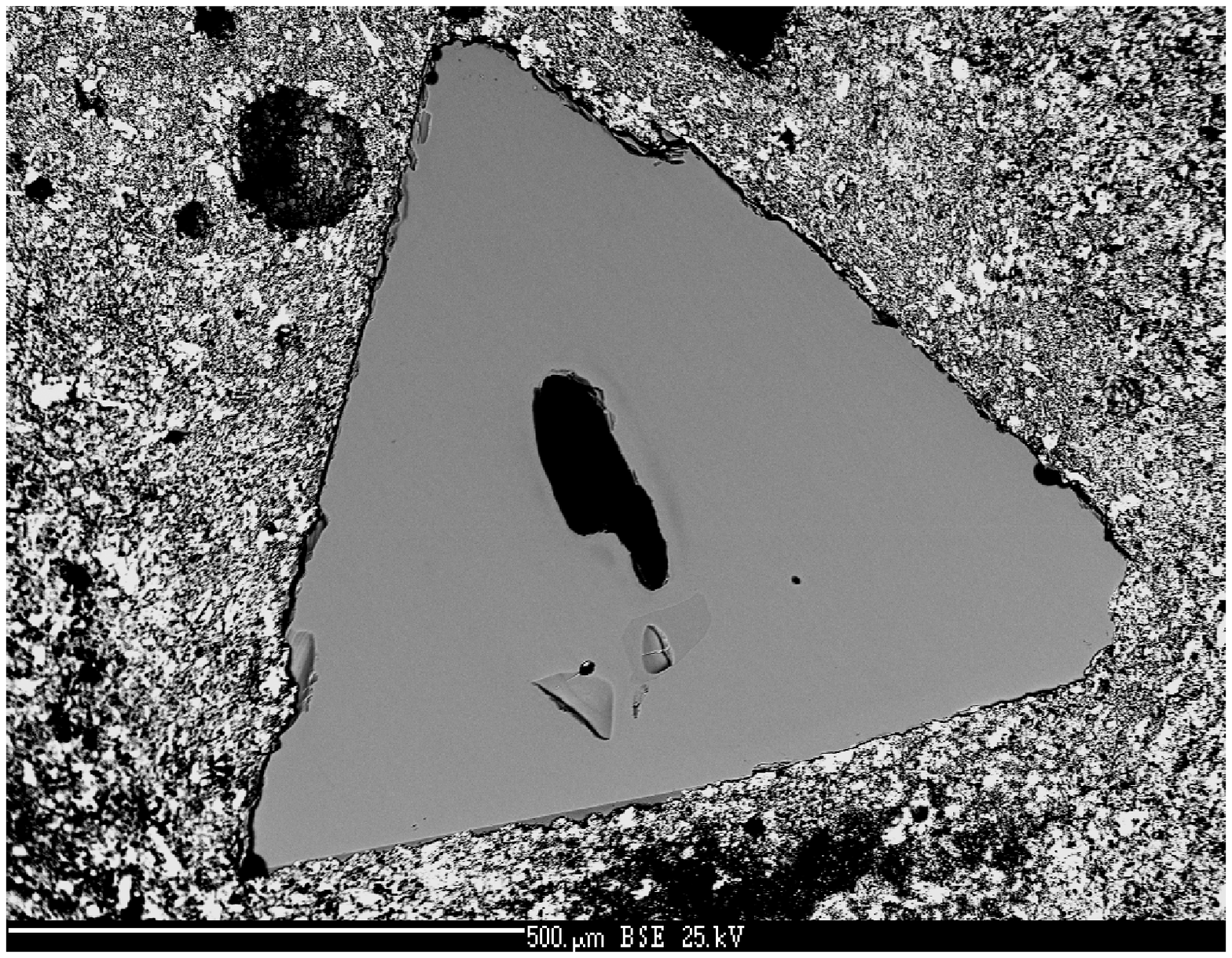}}
\caption{Microstructure of well-facetted $\epsilon$-FeGe-sample
(scanning electron microscope image, BSE contrast) is dominated by
single crystalline area (homogenous grey) and cavity (black) at the
center. A small, disoriented area (darker grey) next to the central
hole shows weak contrast in the BSE image but no significant
differences in composition is found by WDX measurements.}
\label{fig:microprobe}
\end{figure}

The wavelength-dispersive-X-ray-analysis (WDX) performed on a
microprobe (CAMECA SX100, W-cathode) confirms the ratio Fe:Ge = 1:1
of the crystals and X-ray spectra excited by electron beam (25 keV,
20 nA) showed only lines of both elements. The evaluation of the
background intensities at energies of Si-$K$ and I-$L$-lines allowed
us to exclude impurity concentrations larger then 0.08 wt.\% for
both elements involved in the preparation process. Microstructures
prepared from crystals by metallographic methods contained small,
disoriented areas which show unusual strong orientation in the
scanning electron microscope images using back-scattered electron
(BSE) contrast as well as in polarized light of the light microscope
(cf.~Fig.~\ref{fig:microprobe}).

%
%

\section{Single Crystal Structure Determination}
The crystal structure of $\epsilon$-FeGe was solved and refined on
single crystal intensity data (Rigaku R-Axis Spider, Ag-$K{\alpha}$
radiation, $\lambda=56.080$~pm, rotating anode-micro source, with
mirror optic, $\omega$-scan, $\Delta\omega=1.0^\circ$, 160 images;
$2\Theta_{max} = 122.6^\circ$, empirical absorption correction). For
the structure refinement a full-matrix least squares on $F^2$ on
eight parameter was performed using SHELXL-97 \cite{shelxs97}. A
summary of the refinement data is given in
Tab.~\ref{tab:crystallographicdata} and \ref{tab:atomiccoordinates}.
The lattice parameter were determined from powder data on 12
reflections with the program package CSD \cite{Akselrud93} (image
plate Guinier camera HUBER G670, Co-$K\alpha_1$, $\lambda$ =
178.8965 pm, LaB$_6$ ($a$ = 415.692 pm) as internal standard, $T$ =
295 K). All structural parameter are given in
Tab.~\ref{tab:structure}.

%
%
\begin{figure}[t]
\hspace*{-25mm}
\includegraphics[width=0.75\textwidth]{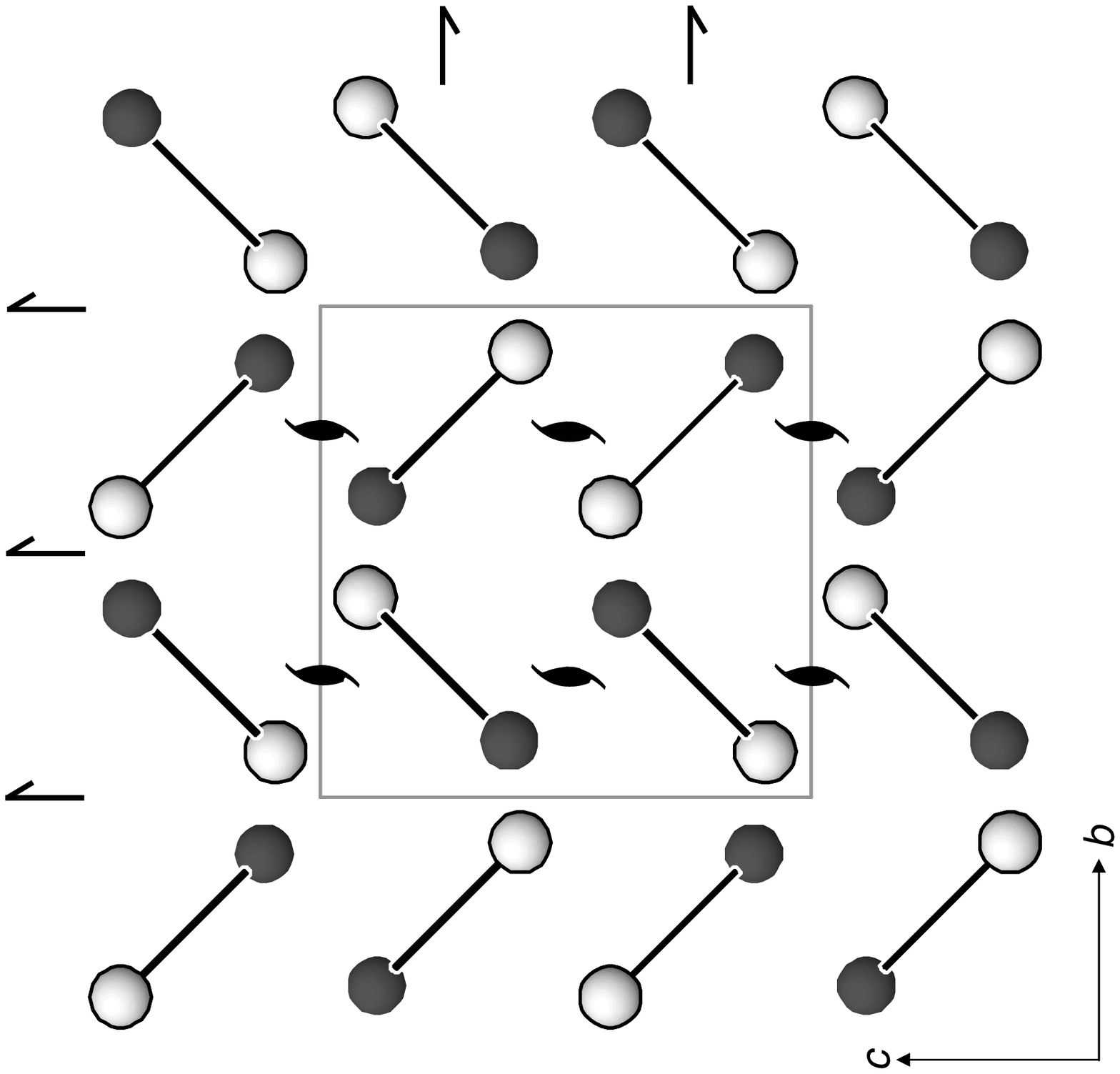}
\caption{Projection of the crystal structure of $\epsilon$-FeGe
along [100]. The symbols show the position of the $2_1$ screw axis.
The lines indicate the shortest Fe-Ge distance.}
\label{fig:projection}
\end{figure}

%
%
\begin{figure}[]
\hspace*{-10mm}
\includegraphics[width=0.575\textwidth]{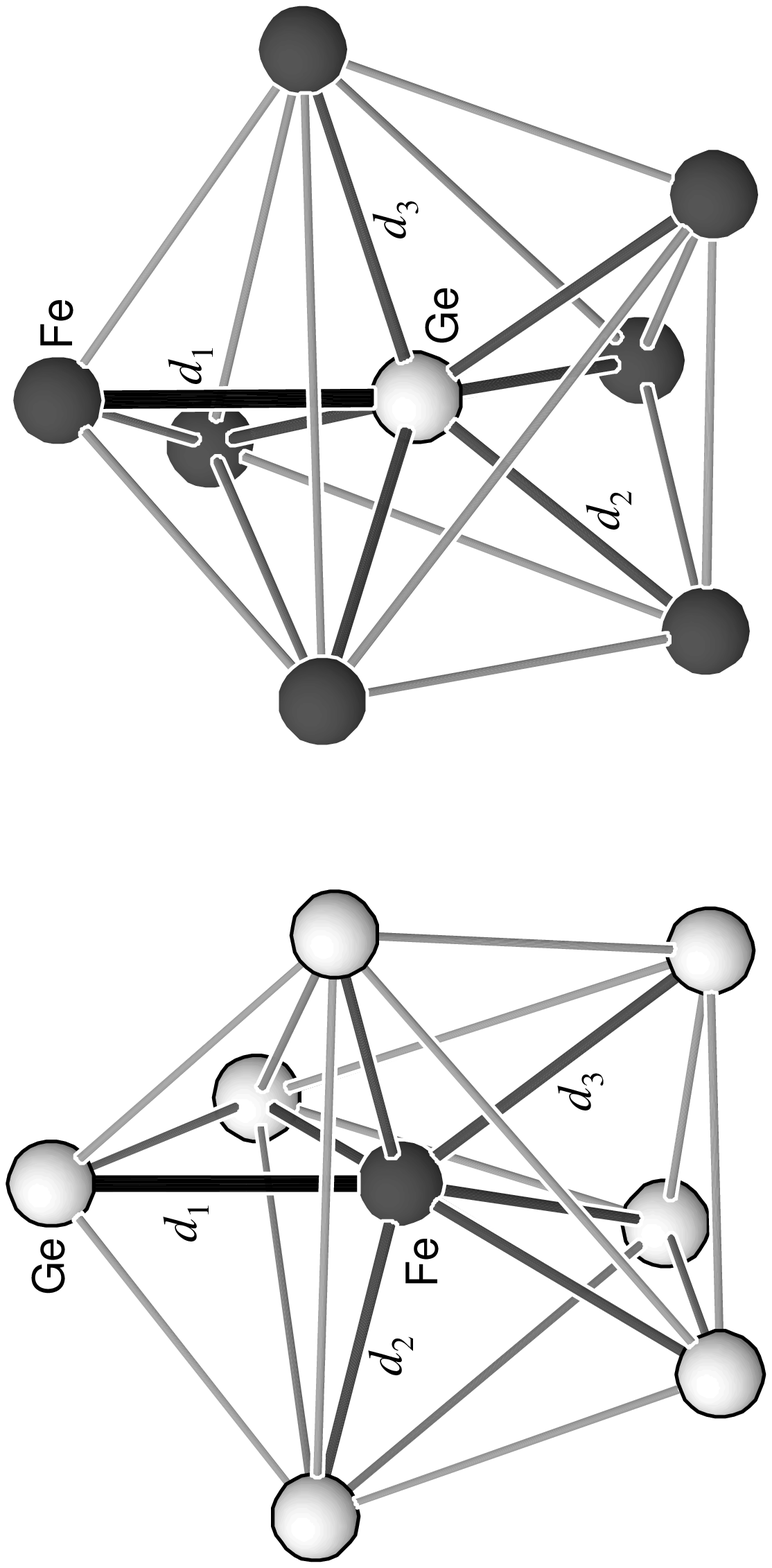}
\caption{Coordination polyhedron of Fe (left) and Ge (right). The
shortest Fe-Ge distance is $d_1$.} \label{fig:coordination}
\end{figure}

\begin{table}[]
\caption{Details of the crystal structure analysis of
$\epsilon$-FeGe at ambient pressure and $T=295$~K.}
\label{tab:crystallographicdata}
\begin{tabular}{l|l}\hline
Density (g cm$^{-3}$)      & 8.22 \\
$\mu$ (cm$^{-1}$)          & 219 \\
measured and unique reflections         & 3680; 1025\\
obs. reflections $> 4.0~\sigma(Fo)$  & 1001\\
$R_{int}$                  & 0.018\\
$R1$; $wR2$              & 0.020; 0.026\\
Goof=S                     & 1.209\\
$\Delta\rho_{max}$;$\Delta\rho_{min}$ (e10$^6$~pm$^3$) & 2.05; -1.01
\\\hline
\end{tabular}
\end{table}

\begin{table}[]
\caption{Anisotropic displacement parameter $U_{ij}$ (in pm$^2$) for
$\epsilon$-FeGe. The $U_{eq}$ is defined as one third of the
orthogonalized $U_{ij}$ tensor.} \label{tab:atomiccoordinates}
\begin{tabular}{l|c|c|c}
   & U$_{11}=U_{22}=U_{33}$ & $U_{23}=U_{13}=U_{12}$ & $U_{eq}$ \\
\hline
Fe   & 57(2)                  & -6(1)    & 57(2) \\ 
Ge   & 55(2)                  &  0(1)    & 55(2) \\ \hline 
\end{tabular}
\end{table}

In the $B20$ structure (space group $P2_13$, No. 198, $Z=4$) both
atoms occupy $4a$ $(x,x,x)$ Wyckoff positions
(Fig.~\ref{fig:projection}). In a compound AB crystallizing in the
ideal $B20$ structure ($x_{\rm{A}}=0.1545$,
$x_{\rm{B}}=1-x_{\rm{A}}$) both atoms have a seven-fold coordination
which lies between that of the two simplest binary cubic structures
NaCl (CN = 6) and CsCl (CN = 8). $\epsilon$-FeGe however, attains a
distorted rock-salt structure where the atoms are slightly shifted
from their ideal position ($x_{\rm{Fe}}=0.13522(2)$ and
$x_{\rm{Ge}}=0.84186(1)$. This results in one short Fe-Ge distance,
$d_1=238.78$~pm, along the trigonal axis pertinent to the Fe atom as
well as two longer ones, $d_2=244.57$~pm, and $d_3=264.45$~pm,
between the three second and three third-nearest neighbors,
respectively (see Fig.~\ref{fig:coordination}). Thus, the
coordination sphere of Fe is build up of seven Ge atoms describing a
strong distorted mono capped trigonal prism.

\section{Powder data at High Pressure}

\begin{table}[]
\caption{Structural parameters of $\epsilon$-FeGe at ambient
conditions and for comparison data of $\epsilon$-FeSi taken from
Ref.~\cite{Wartchow97}. In the standard setting the fractional
coordinates are $x_{\rm{Fe}}=0.38522(2)$, $x_{\rm{Ge}}=0.09186(1)$
for $\epsilon$-FeGe and $x_{\rm{Fe}}=0.38650(2)$,
$x_{\rm{Si}}=0.09262(5)$ for $\epsilon$-FeSi. B denotes either Ge or
Si.} \label{tab:structure}
\begin{tabular}{l|c|c}
                     & $\epsilon$-FeGe  & $\epsilon$-FeSi      \\ \hline  
$a$~(pm)             & 469.95(2)        & 449.5(2)   \\        
$V_0$~(10$^6$pm$^3)$ & 103.79(1)        & 90.8       \\        
$x_{\rm{Fe}}$        & 0.13522(2)       & 0.13650(2) \\        
$x_{\rm B}$          & 0.84186(1)       & 0.84262(5) \\        
$d_1$(pm)            & 238.78(2)        & 228.8      \\        
$d_2$(pm)            & 244.57(1)        & 234.6      \\        
$d_3$(pm)            & 264.45(1)        & 251.98     \\        
$d_{\rm{Fe-Fe}}$(pm) & 288.11(1)        & 275.6      \\        
$d_{\rm{B-B}}$~(pm)  & 291.13(1)        & 278.3      \\ \hline 
\end{tabular}
\end{table}

%
%
\begin{figure}[]
\begin{center}
\includegraphics[width=0.5\textwidth]{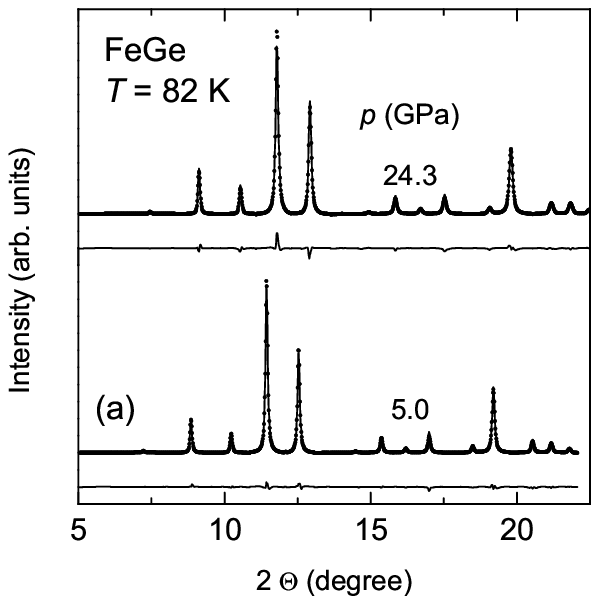}
\includegraphics[width=0.5\textwidth]{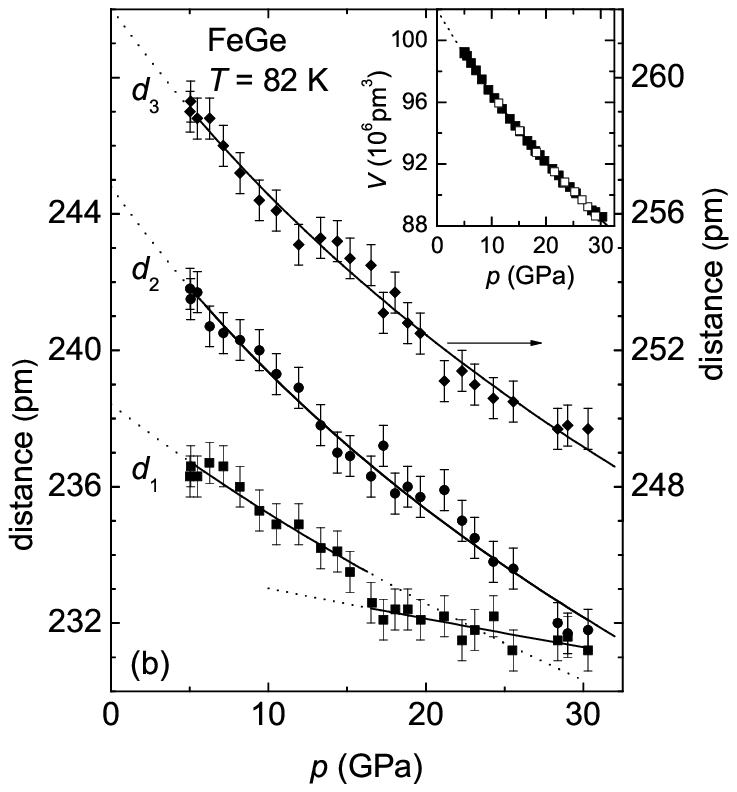}
\caption{(a) Diffraction pattern recorded at 82~K at different
pressures. The difference between raw and fitted data is also shown.
(b) Pressure dependence of the interatomic Fe-Ge distances $d_1$ and
$d_2$ (left scale) and $d_3$ (right scale). The lines represent a
Murnaghan EOS fit to the data. In $d_1(p)$ a discontinuity at about
15~GPa indicates a subtle change in the distances. Error bars
represent $3\sigma$. Inset: $V(p)$ data and EOS fit. Open symbols
represent data obtained on pressure release.}\label{fig:pattern}
\end{center}
\end{figure}

%
%
\begin{figure}[]
\begin{center}
\includegraphics[width=0.5\textwidth]{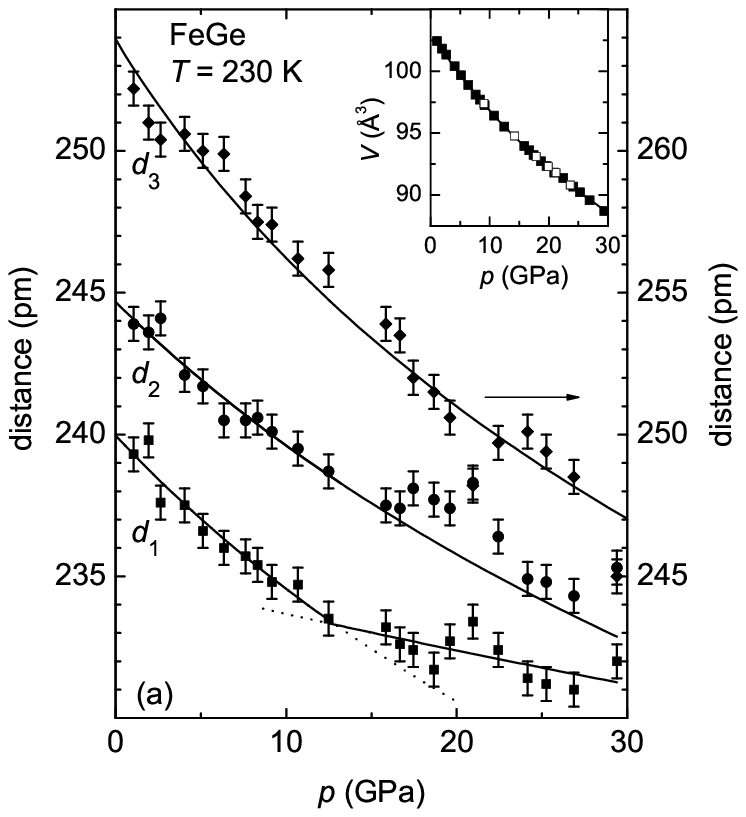}
\includegraphics[width=0.5\textwidth]{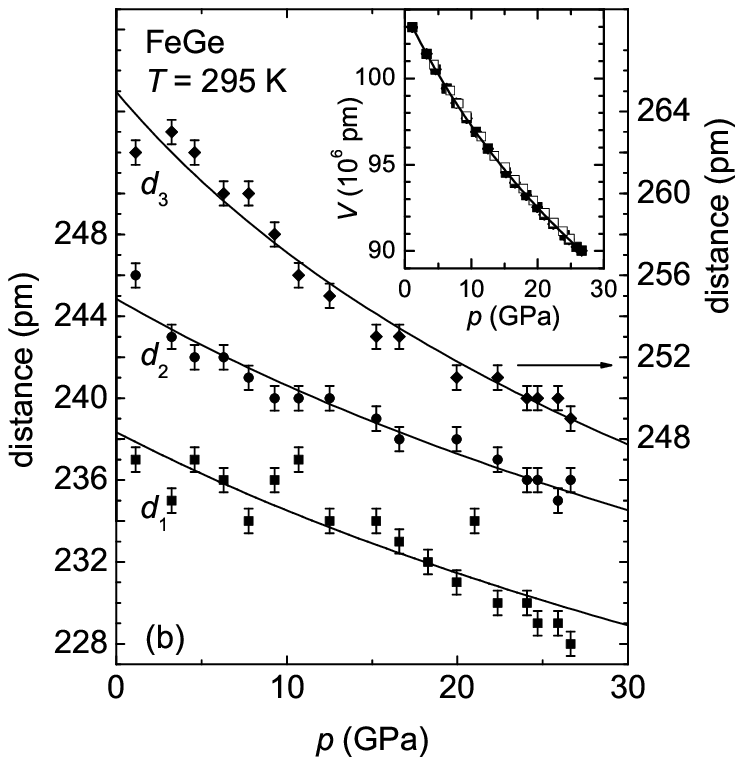}
\caption{Pressure dependence of the interatomic Fe-Ge distances
$d_1$ and $d_2$ (left scale) and $d_3$ (right scale). Error bars
represent $3\sigma$. The lines represent a Murnaghan EOS fit to the
data. (a) $T=230$~K data set, where $d_1(p)$ shows a discontinuity
at about 12~GPa. (b) $T=295$~K data set, where no anomaly in
$d_1(p)$ is observed. Insets: $V(p)$ data and EOS fit. Data obtained
on decreasing pressure are shown as open symbols.}
\label{fig:distanes230}
\end{center}
\end{figure}

The structural evolution with pressure at different temperatures was
determined by angle-dispersive X-ray diffraction experiments at the
beamline ID09 at the European Synchroton Radiation Facility. The
diffraction pattern were collected at a wavelength of
$\lambda=41.254$~pm during a 10~s exposure time with an image plate.
For these measurements well ground and annealed powder (at 670 K for
two days) was loaded in a diamond-anvil cell. Helium was used as
pressure medium and the pressure was determined via the fluorescence
peaks of SrB$_4$O$_7$:Sm$^{2+}$ \cite{Leger90}.

Figure~\ref{fig:pattern}(a) shows two examples of a full profile
structural refinement of the diffraction pattern recorded at 82~K.
The intensities can be very well refined though with little less
precision at the highest pressure. Thus, the fractional atomic
parameter $x_{\rm{Fe}}$ and $x_{\rm{Ge}}$, and hence the interatomic
distances could be deduced. The refinement of the diffraction
pattern measured at 82~K and 230~K revealed clear indications that
the shortest interatomic Fe-Ge distance, $d_1$, which is parallel to
$[111]$, changes its pressure dependence discontinuously
(Fig.~\ref{fig:pattern}(b) and Fig.~\ref{fig:distanes230}(a)). The
extrapolation of the low-pressure $d_1(p)$ behavior for $82\,$K and
$230\,$K described by an appropriate equation-of-state (EOS) clearly
fails to account for the pressure dependence above $15.8\,$GPa and
$12\,$GPa, respectively. These anomalies in $d_1(p)$ agree
remarkably well with the $T_C(p)$-phase boundary deduced from the
electrical resistivity measurements reported in
Ref.~\cite{Pedrazzini07}. The two remaining Fe-Ge distances as well
as all distances at room temperature (Fig.~\ref{fig:distanes230}(b))
decrease smoothly with pressure and no anomaly is seen within the
error bars.

\begin{table}
\caption{Bulk modulus $B_0$, and its pressure derivative $B'$ of
$\epsilon$-FeGe obtained from a fit of the Murnaghan EOS to the
$V(p)$ data. \label{tab:compressibility}}
\begin{tabular}{l|c|c}
T (K) & $B_0$~(GPa) & $B'$ \\ \hline
295   & 130(1)      & 4.7(1) \\
230   & 135(1)      & 4.7(1)  \\
82    & 147(3)      & 4.4(2)  \\
\hline
\end{tabular}
\end{table}

The $V(p)$ dependencies of $\epsilon$-FeGe at all investigated
temperatures revealed no discontinuous pressure dependence and that
the compression is reversible (cf. insets to
Figs.~\ref{fig:pattern}(b) and \ref{fig:distanes230}). All $V(p)$
can be described by a Murnaghan EOS \cite{Murnaghan44}. The
resulting fit parameters are given in
Tab.~\ref{tab:compressibility}. The bulk modulus of $\epsilon$-FeGe
is considerably smaller than that of $\epsilon$-FeSi which is
ranging from $B_0=160(1)$~GPa ($B'=4.0$) \cite{Wood95} to 209(6)~GPa
($B'=5.3$) \cite{Knittle95}. A simple comparison of the unit-cell
volume and using the compressibility of $\epsilon$-FeGe shows that
$\epsilon$-FeGe attains the same unit-cell volume as $\epsilon$-FeSi
at about 25~GPa. Theoretical studies on $\epsilon$-FeSi predict a
phase transformation to the $B2$ structure (CsCl-type) at about
13~GPa \cite{Vocadlo99} or in the range of 30 to 40~GPa
\cite{Caracas04}. On the other hand, no phase transformation in
$\epsilon$-FeSi has been found up to 50~GPa \cite{Knittle95}. Thus,
more structural information at pressures well above 30~GPa is needed
to resolve the stability of $\epsilon$-FeGe.

We acknowledge W. Schnelle for characterizing the samples by
magnetic measurements and H. Borrmann for single crystal data
collection. We are grateful to R. Demchnya and A. Wosylus for their
commitment during the high pressure X-ray experiments.

\end{document}